\documentclass[12pt]{article}
\usepackage{amsmath}
\usepackage{color}
\usepackage{graphicx}
\begin{document}
\baselineskip=18 pt
\begin{center}
{\large{\bf Spin-$0$ scalar particle interacts with scalar potential in the presence of magnetic field and quantum flux under the effects of KKT in $5D$ cosmic string space-time  }}
\end{center}

\vspace{.5cm}

\begin{center}
{\bf Faizuddin Ahmed}\footnote{ \bf faizuddinahmed15@gmail.com ; faiz4U.enter@rediffmail.com}\\ 
{\bf Maryam Ajmal Women's College of Science \& Technology, Hojai-782435, Assam, India}
\end{center}

\vspace{.5cm}

\begin{abstract}

In this paper, we study a relativistic quantum dynamics of spin-$0$ scalar particle interacts with scalar potential in the presence of a uniform magnetic field and quantum flux in background of Kaluza-Klein theory (KKT). We solve Klein-Gordon equation in the considered framework and analyze the relativistic analogue of the Aharonov-Bohm effect for bound states. We show that the energy levels depend on the global parameters characterizing the space-time, scalar potential and the magnetic field which break their degeneracy. 

\end{abstract}

{\bf keywords:} cosmic string space-time, Relativistic wave equation, electromagnetic interactions, energy spectrum, wave-functions, Aharonov-Bohm effect, special functions.

\vspace{0.1cm}


\section{Introduction}

The relativistic wave-equations are of current research interest for theoretical physicists \cite{BT,AWT} including those in nuclear and high energy physics \cite{TYW,WG}. In recent years, many studies have been carried out to explore the relativistic energy eigenvalues and eigenfunctions in the background of cosmic string space-times ({\it e. g.}, \cite{BCL,HC,SZ,EAFB,KBB,MHHH,MM,MHHH2,MHHH3,EPL}).

Cosmic strings have been produced by phase transition in the early universe \cite{TWBK,AV} as predicted in the string theory \cite{MH2} as well as in the particle physics \cite{AV3,MH3}. These include domain walls, cosmic string and monopoles. Among them, cosmic string and monopoles are the best candidates to be observed. Cosmic string \cite{AV2} and global monopoles \cite{MB} are exotic topological objects that modify the space-time geometry, producing a planar and solid angle deficit, respectively. A series of cylindrically symmetric solutions of the Einstein and Einstein-Gauss-Bonnet equations in the context of Kaluza-Klein theory was investigated in \cite{MAA}.

In this paper, we study KG-equation in the background of $5D$ cosmic string space-time geometry produced by topological defects in the context of Kaluza-Klein theory \cite{Th,OZK,TM,DB}. Topological defects play an important role in condensed matter physics systems \cite{MB,AMdM,CS,CS2,CS3,MJB,LD,WCFS,PRD,SZZ} where, topological defects analogue to cosmic string appear in phase transitions in liquid crystals \cite{HM,FM}. Geometric quantum phases \cite{AB,JGA,JRN,KBCF,CALR} is a topological defect phenomena that describe phase shifts acquire by the wave-function of a quantum mechanical particle. A well-known quantum phase is the Aharonov-Bohm effect \cite{MP,VBB,YA} which arises due to the presence of magnetic quantum flux produced by topological defects space-times. This effect has investigated in, Newtonian theory \cite{MAA2}, bound states of massive fermions \cite{VRK}, scattering of dislocated wave-fronts \cite{CC}, position-dependent mass system under the effects of torsion \cite{IJMPD,AHEP,R1,R2,R3,AHEP3,CJP}, bound states of spin-$0$ scalar particle \cite{EPJP,IJMPA,MPLA,IJGMMP}. In the context of Kaluza-Klein theory, this effect has studied by many authors, such as, spin-$0$ scalar particle \cite{CF,CF2,EVBL,EVBL2,EVBL3,EPJC,AHEP2,JC,EVBL4}, under torsion effects \cite{GG,GG2,IVV,DEN,MWK,CHO,KHS}, fermions \cite{DEN,YSW,AM,SI}, Lorentz symmetry violation \cite{SMC,MG,APB}, and in graphene layer \cite{KB}.

\section{Spin-$0$ scalar particle interacts with potential in $5D$ cosmic string geometry within the KKT }

In the context of Kaluza-Klein theory \cite{Th,OZK}, the metric with a magnetic quantum flux ($\Phi$) passing along the symmetry axis of the string assumes the following form
\begin{equation}
ds^2=-dt^2+dr^2+\alpha^2\,r^2\,d\phi^2+dz^2+\left[ dx+K\,A_{\mu} ({\bf x})\, dx^{\mu} \right ]^2,
\label{1}
\end{equation}
where $t$ is the time-coordinate, $x$ is the coordinate associated with fifth additional dimension having ranges $0 < x < 2\,\pi\,a$ where, $a$ is the radius of the compact dimension of $x$, $(r, \phi, z)$ are the cylindrical coordinates with the usual ranges, and $K$ is the Kaluza constant \cite{CF}. The parameter $\alpha=(1-4\,\mu)$ \cite{AV} characterizing the wedge parameter which determines the angular deﬁcit, $\nabla\,\phi=2\,\pi\,(1-\alpha)$ and $\mu$ is the linear mass density of the string. The cosmology and gravitation imposes limits to the range of the wedge parameter $\alpha$ with $0 < \alpha < 1$ \cite{MOK,PLA}.

Based on \cite{CF,EVBL,AHEP2,JC}, we introduce a uniform magnetic field $B_0$ and Aharonov-Bohm magnetic flux $\Phi$ \cite{MP} through the line-element of $5D$ cosmic string space-time geometry (\ref{1}) in the following form
\begin{equation}
ds^2=-dt^2+dr^2+\alpha^2\,r^2\,d\phi^2+dz^2+\left[ dx+\left(-\frac{1}{2}\,\alpha\,B_0\,r^2+\frac{\Phi}{2\pi} \right)\, d\phi \right ]^2,
\label{2}
\end{equation}
where the gauge configuration given by
\begin{equation}
A_{\phi}=K^{-1}\,\left (-\frac{1}{2}\,\alpha\,B_{0}\,r^2+\frac{\Phi}{2\pi} \right)
\label{3}
\end{equation}
gives rise to a uniform magnetic field $\vec{B}= \vec{\nabla}\times \vec{A}=-K^{-1}\,B_0\,\hat{z}$ \cite{GAM}, $\hat{z}$ is the unitary vector in the $z$-direction. Here $\Phi=const$ is the magnetic quantum flux \cite{YA,MP,GAM} through the core of the topological defects \cite{CF3}.

The relativistic quantum dynamics of spin-$0$ scalar particle interacts with scalar potential $S (r)$ by modifying the mass term $ m \rightarrow m+ S (r)$ \cite{WG} is described by the following equation \cite{IJMPD,AHEP,AHEP2,EPJP,EPJC}:
\begin{equation}
\left [\frac{1}{\sqrt{-g}}\,\partial_{\mu} (\sqrt{-g}\,g^{\mu\nu}\,\partial_{\nu})-(m+S)^2 \right]\,\Psi=0,
\label{4}
\end{equation}
with $g$ is the determinant of metric tensor with $g^{\mu\nu}$ its inverse and $m$ is rest mass of the scalar particle. The scalar potential $S(r)$ is usually observed under the static field conditions. 

For the metric (\ref{2}) the metric tensor $g^{\mu\nu}$ is 
\begin{equation}
g^{\mu\nu}=\left (\begin{array}{lllll}
-1 & 0 & \quad 0 & 0 & \quad 0 \\
\quad 0 & 1 & \quad 0 & 0 & \quad 0 \\
\quad 0 & 0 & \quad \frac{1}{\alpha^2\,r^2} & 0 & -\frac{K\,A_{\phi}}{\alpha^2\,r^2} \\
\quad 0 & 0 & \quad 0 & 1 & \quad 0 \\
\quad 0 & 0 & -\frac{K\,A_{\phi}}{\alpha^2\,r^2} & 0 & 1+\frac{K^2\,A^2_{\phi}}{\alpha^2\,r^2}
\end{array} \right).
\label{5}
\end{equation}

By considering the line-element (\ref{2}) into the Eq. (\ref{4}), we obtain the following differential equation :
\begin{eqnarray}
&&[-\frac{\partial^2}{\partial t^2}+\frac{\partial^2}{\partial r^2}+\frac{1}{r}\,\frac{\partial}{\partial r}+\frac{1}{\alpha^2\,r^2}\,\left(\frac{\partial}{\partial \phi}-K\,A_{\phi}\,\frac{\partial}{\partial x}\right)^2+\frac{\partial^2}{\partial z^2}+\frac{\partial^2}{\partial x^2}\nonumber\\
&&-\left (m+S \right)^2]\,\Psi (t,r,\phi,z)=0.
\label{6}
\end{eqnarray}
Since the line-element (\ref{2}) is independent of $t, \phi ,z, x$. One can choose the following ansatz for the function $\Psi$ as:
\begin{equation}
\Psi(t, r, \phi, z, x)=e^{i\,(-E\,t+l\,\phi+k\,z+q\,x)}\,\psi(r),
\label{7}
\end{equation}
where $E$ is the energy of the particle, $l=0,\pm\,1,\pm\,2,.. \in {\bf Z}$, and $k, q$ are constants.

Substituting the ansatz (\ref{7}) into the Eq. (\ref{6}), we obtain the following equation:
\begin{equation}
\left [\frac{d^2}{dr^2}+\frac{1}{r}\,\frac{d}{dr}+E^2-k^2-q^2-\frac{(l-K\,q\,A_{\phi})^2}{\alpha^2\,r^2}-\left (m+S \right)^2 \right]\,\psi (r)=0.
\label{8}
\end{equation}

Below we consider interactions of scalar particle with potential of different kind, such as, Cornell-type, Coulomb-type and linear confining potential and evaluate the energy eigenvalues and eigenfunctions in details. In addition, we consider zero scalar potential in the considered relativistic system and obtain the energy eigenvalues and analyze the effects of topological defects as well as the magnetic field.

\vspace{0.5cm}
{\bf Case A\quad:\quad Interactions with Cornell-type potential}
\vspace{0.5cm}

Cornell-type potential consists of linear plus Coulomb-like term is a particular case of the quark-antiquark interaction \cite{MKB,CA}. The Coulomb potential is responsible at small distances or short range interactions and linear potential leads to confinement of quark. This type of potential is of the form \cite{ZW,AHEP,AHEP3} 
\begin{equation}
S (r)=\frac{\eta_c}{r}+\eta_L\,r
\label{9}
\end{equation}
where $\eta_c, \eta_L$ are the potential parameters.

Substituting Eqs. (\ref{3}) and (\ref{9}) into the Eq. (\ref{8}), we obtain the following equation:
\begin{equation}
\left [\frac{d^2}{dr^2}+\frac{1}{r}\,\frac{d}{dr}+\lambda-\frac{j^2}{r^2}-\Omega^2\,r^2-\frac{a}{r}-b\,r \right]\,\psi (r)=0,
\label{10}
\end{equation}
where
\begin{eqnarray}
\lambda&=&E^2-k^2-q^2-m^2-2\,\eta_c\,\eta_L-2\,m\,\omega\,\frac{(l-\frac{q\,\Phi}{2\pi})}{\alpha},\nonumber\\
\Omega&=&\sqrt{m^2\,\omega^2+\eta^2_{L}},\nonumber\\
j&=&\sqrt{\frac{(l-\frac{q\,\Phi}{2\pi})^2}{\alpha^2}+\eta^2_{c}},\nonumber\\
\omega&=&\frac{q\,B_0}{2\,m},\nonumber\\
a&=&2\,m\,\eta_c,\nonumber\\
b&=&2\,m\,\eta_L.
\label{11}
\end{eqnarray}

Introducing a new variable $\rho=\sqrt{\Omega}\,r$, Eq. (\ref{10}) becomes
\begin{equation}
\left [\frac{d^2}{d\rho^2}+\frac{1}{\rho}\,\frac{d}{d\rho}+\zeta-\frac{j^2}{\rho^2}-\rho^2-\frac{\eta}{\rho}-\theta\,\rho \right]\,\psi (\rho)=0,
\label{12}
\end{equation}
where
\begin{equation}
\zeta=\frac{\lambda}{\Omega}\quad,\quad \eta=\frac{a}{\sqrt{\Omega}}\quad,\quad \theta=\frac{b}{\Omega^{\frac{3}{2}}}.
\label{13}
\end{equation}

Suppose the possible solution to Eq. (\ref{12}) is
\begin{equation}
\psi (\rho)=\rho^{j}\,e^{-\frac{1}{2}\,(\rho+\theta)\,\rho}\,H (\rho).
\label{14}
\end{equation}
Substituting the solution Eq. (\ref{14}) into the Eq. (\ref{12}), we obtain
\begin{equation}
H''(\rho)+\left [\frac{\gamma}{\rho}-\theta-2\,\rho \right ]\,H'(\rho)+\left [-\frac{\beta}{\rho}+\Theta \right]\,H (\rho)=0,
\label{15}
\end{equation}
where
\begin{eqnarray}
&&\gamma=1+2\,j,\nonumber\\
&&\Theta=\zeta+\frac{\theta^2}{4}-2\,(1+j),\nonumber\\
&&\beta=\eta+\frac{\theta}{2}\,(1+2\,j).
\label{16}
\end{eqnarray}
Equation (\ref{15}) is the biconfluent Heun's differential equation \cite{IJMPD,AHEP,EPJC,AHEP2,AR,SYS,ERFM, KBCF2} and $H (\rho)$ is the Heun polynomials.

The above equation (\ref{15}) can be solved by the Frobenius method. We consider the power series solution \cite{GBA}
\begin{equation}
H (\rho)=\sum_{i=0}^{\infty}\,c_{i}\,\rho^{i}
\label{17}
\end{equation}
Substituting the above power series solution into the Eq. (\ref{15}), we obtain the following recurrence relation for the coefficients:
\begin{equation}
c_{n+2}=\frac{1}{(n+2)(n+2+2\,j)}\,\left[\left\{\beta+\theta\,(n+1) \right\}\,c_{n+1}-(\Theta-2\,n)\,c_{n} \right].
\label{18}
\end{equation}
And the various coefficients are
\begin{eqnarray}
&&c_1=\left(\frac{\eta}{\gamma}+\frac{\theta}{2} \right)\,c_0,\nonumber\\
&&c_2=\frac{1}{4\,(1+j)}\,[\left(\beta+\theta \right)\,c_{1}-\Theta\,c_{0}].
\label{19}
\end{eqnarray}

We must truncate the power series by imposing the following two conditions \cite{IJMPD,AHEP,EVBL,EVBL2,EVBL3, EPJP,EPJC,AHEP2,ERFM,KBCF2}:
\begin{eqnarray}
\Theta&=&2\,n, \quad (n=1,2,...)\nonumber\\
c_{n+1}&=&0.
\label{21}
\end{eqnarray} 

By analyzing the condition $\Theta=2\,n$, we get the following second degree expression of the energy eigenvalues $E_{n,l}$:
\begin{eqnarray}
&&\frac{\lambda}{\Omega}+\frac{\theta^2}{4}-2\,(1+j)=2\,n\nonumber\\\Rightarrow
&&E_{n,l}=\pm\,\{k^2+q^2+m^2+2\,\Omega\,\left(n+1+\sqrt{\frac{(l-\frac{q\,\Phi}{2\pi})^2}{\alpha^2}+\eta^2_{c}} \right)\nonumber\\
&&+2\,\eta_c\,\eta_L+2\,m\,\omega\,\frac{(l-\frac{q\,\Phi}{2\pi})}{\alpha}-\frac{m^2\,\eta^2_{L}}{\Omega^2} \}^{\frac{1}{2}}.
\label{22}
\end{eqnarray}
For $\alpha \rightarrow 1$, the relativistic energy eigenvalue (\ref{22}) is consistent with those result in \cite{EVBL2}.

Now, we impose additional recurrence condition $c_{n+1}=0$ to find the individual energy levels and wave-functions one by one as done in \cite{IJMPD,AHEP,EPJP,EPJC,AHEP2,ERFM}. For $n=1$, we have $\Theta=2$ and $c_2=0$ which implies from Eq. (\ref{19})
\begin{eqnarray}
&&c_1=\frac{2}{\beta+\theta}\,c_0\Rightarrow\left(\frac{\eta}{1+2\,j}+\frac{\theta}{2} \right)=\frac{2}{\beta+\theta}\nonumber\\
&&\Omega^3_{1,l}-\frac{a^2}{2\,(1+2\,j)}\,\Omega^2_{1,l}-a\,b\,(\frac{1+j}{1+2\,j})\,\Omega_{1,l}-\frac{b^2}{8}\,(3+2\,j)=0
\label{23}
\end{eqnarray}
a constraint on the parameter $\Omega_{1,l}$. The magnetic field $B^{1,l}_{0}$ is so adjusted that Eq. (\ref{23}) can be satisfied and we have simplified by labelling as:
\begin{equation}
\omega_{1,l}=\frac{1}{m}\sqrt{\Omega^2_{1,l}-\eta^2_{L}} \leftrightarrow B^{1,l}_{0}=\frac{2}{q}\sqrt{\Omega^2_{1,l}-\eta^2_{L}}.
\label{24}
\end{equation}

Therefore, the ground state energy level for $n=1$ is given by
\begin{eqnarray}
&&E_{1,l}=\pm\,\{k^2+q^2+m^2+2\,\Omega_{1,l}\,\left(2+\sqrt{\frac{(l-\frac{q\,\Phi}{2\pi})^2}{\alpha^2}+\eta^2_{c}} \right)\nonumber\\
&&+2\,\eta_c\,\eta_L+2\,m\,\omega_{1,l}\,\frac{(l-\frac{q\,\Phi}{2\pi})}{\alpha}-\frac{m^2\,\eta^2_{L}}{\Omega^2_{1,l}}\}^{\frac{1}{2}}.
\label{25}
\end{eqnarray}

Thus we have seen that the presence of Cornell-type potential allow the formation of bound states solution of the system and the cosmic string parameter $\alpha$ modified the energy profile for each radial mode. In addition, the energy level get shifted due to the quantum flux $\Phi$ present in the relativistic system which gives us an analogous of the Aharonov-Bohm effect making them a periodic function with periodicity $\Phi_0=\frac{2\,\pi}{q}\,\tau$, with $\tau=0,1,2,...$, that is, $E_{n, l} (\Phi+\Phi_0)=E_{n, l \mp \tau} (\Phi)$.

The radial wave-functions is 
\begin{eqnarray}
\psi_{1,l}=\rho^{\sqrt{\frac{(l-\frac{q\,\Phi}{2\pi})^2}{\alpha^2}+\eta^2_{c}}}\,e^{-\frac{1}{2}\,\left (\frac{2\,m\,\eta_L}{\Omega^{\frac{3}{2}}_{1,l}}+\rho \right)\,\rho}\,\left(c_0+c_1\,\rho\right),
\label{26}
\end{eqnarray}
where
\begin{eqnarray}
c_1&=&\left(\frac{2\,m\,\eta_c}{\sqrt{\Omega_{1,l}}\,(1+2\,\sqrt{\frac{(l-\frac{q\,\Phi}{2\pi})^2}{\alpha^2}+\eta^2_{c}})}+\frac{m\,\eta_L}{\Omega^{\frac{3}{2}}_{1,l}} \right)\,c_0.
\label{27}
\end{eqnarray}

\vspace{0.5cm}
{\bf Case B\quad:\quad Interactions with Coulomb-type potential }
\vspace{0.5cm}

We consider $\eta_L \rightarrow 0$ in the scalar potential $S (r)$ considered earlier. Thus the Coulomb-type potential is of the form 
\begin{equation}
S (r)=\frac{\eta_c}{r},
\label{29}
\end{equation}
This kind of potential has used to study the position-dependent mass systems \cite{AHEP2,EVBL2,ALCO,MSC} in the relativistic quantum mechanics.

The radial wave-equations Eq. (\ref{10}) becomes
\begin{equation}
\left [\frac{d^2}{dr^2}+\frac{1}{r}\,\frac{d}{dr}+\tilde{\lambda}-\frac{j^2}{r^2}-m^2\,\omega^2\,r^2-\frac{a}{r} \right]\,\psi (r)=0,
\label{30}
\end{equation}
where $\tilde{\lambda}=E^2-k^2-q^2-m^2-2\,m\,\omega\,\frac{(l-\frac{q\,\Phi}{2\pi})}{\alpha}$. 

Introduce a new variable $\rho=\sqrt{m\,\omega}\,r$, Eq. (\ref{30}) becomes
\begin{equation}
\left [\frac{d^2}{d\rho^2}+\frac{1}{\rho}\,\frac{d}{d\rho}+\frac{\tilde{\lambda}}{m\,\omega}-\frac{j^2}{\rho^2}-\rho^2-\frac{\tilde{\eta}}{\rho} \right]\,\psi (\rho)=0,
\label{31}
\end{equation}
where $\tilde{\eta}=\frac{a}{\sqrt{m\,\omega}}$.

Suppose the possible solution to Eq. (\ref{31}) is
\begin{equation}
\psi (\rho)=\rho^{j}\,e^{-\frac{\rho^2}{2}}\,H (\rho).
\label{32}
\end{equation}
Substituting the solution Eq. (\ref{14}) into the Eq. (\ref{12}), we obtain
\begin{equation}
H''(\rho)+\left [\frac{1+2\,j}{\rho}-2\,\rho \right ]\,H'(\rho)+\left [-\frac{\tilde{\eta}}{\rho}+\tilde{\Theta} \right]\,H (\rho)=0,
\label{33}
\end{equation}
where $\tilde{\Theta}=\frac{\tilde{\lambda}}{m\,\omega}-2\,(1+j)$.

Equation (\ref{33}) is the biconfluent Heun's differential equation \cite{IJMPD,AHEP,EPJP,EPJC,AHEP2,AR,SYS, ERFM,KBCF2} and $H (\rho)$ is the Heun polynomials.

Substituting the power series solution (\ref{17}) into the Eq. (\ref{33}), we obtain the following recurrence relation for the coefficients:
\begin{equation}
c_{n+2}=\frac{1}{(n+2)(n+2+2\,j)}\,\left[ \tilde{\eta}\,c_{n+1}-(\tilde{\Theta}-2\,n)\,c_{n} \right].
\label{34}
\end{equation}
And the various coefficients are
\begin{equation}
c_1=\frac{\tilde{\eta}}{1+2\,j}\,c_0\quad,\quad c_2=\frac{1}{4\,(1+j)}\,[\tilde{\eta}\,c_{1}-\Theta\,c_{0}].
\label{35}
\end{equation}
The power series expansion (\ref{17}) becomes a polynomial of degree $n$ by imposing two conditions \cite{IJMPD,AHEP,EVBL,EVBL2,EVBL3,EPJP,EPJC,AHEP2,ERFM,KBCF2}:
\begin{equation}
c_{n+1}=0\quad,\quad \tilde{\Theta}=2\,n \quad (n=1,2,....) 
\label{36}
\end{equation}

By analyzing the condition $\tilde{\Theta}=2\,n$, we get the following energy eigenvalues $E_{n,l}$:
\begin{equation}
E_{n,l}=\pm\,\sqrt{k^2+q^2+m^2+2\,m\,\omega\,\left(n+1+\sqrt{\frac{(l-\frac{q\,\Phi}{2\pi})^2}{\alpha^2}+\eta^2_{c}}+\frac{(l-\frac{q\,\Phi}{2\pi})}{\alpha} \right)}.
\label{37}
\end{equation}

For the radial mode $n=1$, we have $\tilde{\Theta}=2$ and $c_2=0$ which implies
\begin{equation}
\omega_{1,l}=\frac{2\,m\,\eta^2_{c}}{\left(1+2\,\sqrt{\frac{(l-\frac{q\,\Phi}{2\pi})^2}{\alpha^2}+\eta^2_{c}}\right)}\leftrightarrow B^{1,l}_{0}=\frac{4\,m^2\,\eta^2_{c}}{q\,\left(1+2\,\sqrt{\frac{(l-\frac{q\,\Phi}{2\pi})^2}{\alpha^2}+\eta^2_{c}}\right)}.
\label{38}
\end{equation}
a constraint on the parameter $\omega_{1,l}$ or the magnetic field $B^{1,l}_0$.

The ground state energy eigenvalues for $n=1$ is
\begin{equation}
E_{1,l}=\pm\,\sqrt{k^2+q^2+m^2+2\,m\,\omega_{1,l}\,\left(2+\sqrt{\frac{(l-\frac{q\,\Phi}{2\pi})^2}{\alpha^2}+\eta^2_{c}}+\frac{(l-\frac{q\,\Phi}{2\pi})}{\alpha} \right)},
\label{39}
\end{equation}
where $\omega_{1,l}$ is given by Eq. (\ref{38}).

Equation (\ref{39}) corresponds to the allowed values of the energy level for the radial mode $n=1$ of the system in the context of Kaluza-Klein theory. For $\alpha \rightarrow 1$, the energy eigenvalues (\ref{37}) is consistent with those result in \cite{EVBL2}. Since the values of the wedge parameter $\alpha$ lies in the range $\alpha <1$, hence, the energy eigenvalues presented here shifted in comparison to those result in \cite{EVBL2} and break their degeneracy. We observed here that the magnetic field $B_0$ depends on the quantum numbers $\{l,n\}$ of the system with their possible values are determined by a relation. The energy level get shifted due to the presence of quantum flux $\Phi$ in the system which gives us an analogous of the Aharonov-Bohm effect making them a periodic function with periodicity $\Phi_0=\frac{2\,\pi}{q}\,\tau$, with $\tau=0,1,2,...$, that is, $E_{n, l} (\Phi+\Phi_0)=E_{n, l \mp \tau} (\Phi)$.

The ground state wave-function ($n=1$) is given by
\begin{equation}
    \psi_{1,l}=e^{-\frac{\rho^2}{2}}\,\rho^{\sqrt{\frac{(l-\frac{q\,\Phi}{2\pi})^2}{\alpha^2}+\eta^2_{c}}}\,(c_0+c_1\,\rho),
    \label{400}
\end{equation}
where 
\begin{equation}
    c_1=\frac{1}{\sqrt{\frac{1}{2}+\sqrt{\frac{(l-\frac{q\,\Phi}{2\pi})^2}{\alpha^2}+\eta^2_{c}}}}.
    \label{410}
\end{equation}

\vspace{0.5cm}
{\bf Special case}
\vspace{0.5cm}

In this special case, we choose zero magnetic field, $B_0 \rightarrow 0$ into the system. The radial wave-equations Eq. (\ref{30}) becomes
\begin{equation}
\left [\frac{d^2}{dr^2}+\frac{1}{r}\,\frac{d}{dr}+E^2-k^2-q^2-m^2-\frac{j^2}{r^2}-\frac{a}{r} \right]\,\psi (r)=0.
\label{bb1}
\end{equation} 
The above can now be expressed as \cite{IJMPA,AFN}
\begin{equation}
\psi ''(r)+\frac{1}{r}\,\psi' (r)+\frac{1}{r^2}\,(-\xi_1\,r^2+\xi_2\,r-\xi_3)\,\psi (r)=0.
\label{bb2}
\end{equation} 
where
\begin{equation}
\xi_1=k^2+q^2+m^2-E^2\quad,\quad \xi_2=-a\quad,\quad \xi_3=j^2.
\label{bb3}
\end{equation}

The energy eigenvalues is given by
\begin{equation}
E_{n',l}=\pm\,m\,\sqrt{1-\frac{\eta^2_{c}}{\left(n'+\frac{1}{2}+\sqrt{\frac{(l-\frac{q\,\Phi}{2\pi})^2}{\alpha^2}+\eta^2_{c}} \right)^2}+\frac{k^2}{m^2}+\frac{q^2}{m^2}},
\label{bb4}
\end{equation}
where $n'=0,1,2,....$. 

Equation (\ref{bb4}) is the energy eigenvalues of a scalar particle in cosmic string background within the Kaluza-Klein theory without magnetic field. For $\alpha \rightarrow 1$, the energy eigenvalues Eq. (\ref{bb4}) is consistent with those result in \cite{EVBL}. Since the values of the wedge parameter $\alpha$ lies in the range $\alpha <1$, hence, the energy levels presented here shifted in comparison to those result in \cite{EVBL} and break their degeneracy. The presence of quantum flux $\Phi$ in the relativistic system shifted the energy levels and the relativistic analogue of the Aharonov-Bohm effect is observed by making them a periodic function with periodicity $\Phi_0=\frac{2\,\pi}{q}\,\tau$, with $\tau=0,1,2,...$, that is, $E_{n', l} (\Phi+\Phi_0)=E_{n', l \mp \tau} (\Phi)$.

The corresponding radial wave functions is given by
\begin{equation}
\psi_{n',l} (r)=|N|_{n',l}\,r^{j}\,e^{-\sqrt{k^2+q^2+m^2-E^2_{n',l}}\,r}\,L^{(2\,j)}_{n'} (r),
\label{bb5}
\end{equation}
where $|N|_{n',l}=2^{2\,j}\,\left(k^2+q^2+m^2-E^2_{n',l} \right)^{j+\frac{1}{2}}\,\left(\frac{n'!}{(n'+2\,j)!}\right)^{\frac{1}{2}}$ is the normalization constant and $L^{(2\,j)}_{n'} (r)$ is the generalized Laguerre polynomials. The polynomials $L^{(j)}_{n'} (r)$ are orthogonal over $[0,\infty)$ with respect to the measure with weighting function $r^{j}\,e^{-r}$ as
\begin{equation}
\int^{\infty}_{0} r^{j}\,e^{-r}\,L^{(j)}_{n'}\,L^{(j)}_{m'}\,dr=\frac{(n'+j)!}{n'!}\,\delta_{n'\,m'}.
\label{int}
\end{equation}

\vspace{0.5cm}
{\bf Case C\quad:\quad Interactions with Linear potential}
\vspace{0.5cm}

We consider $\eta_c \rightarrow 0$ in the potential $S (r)$ considered earlier. Thus the linear scalar potential is of the form 
\begin{equation}
S (r)=\eta_L\,r.
\label{40}
\end{equation} 
The linear potential have studied by many authors in the relativistic quantum mechanics \cite{ERFM,ALCO,MSC, R4,R5,R6}.

The radial wave-equations Eq. (\ref{10}) becomes
\begin{equation}
\left [\frac{d^2}{dr^2}+\frac{1}{r}\,\frac{d}{dr}+\tilde{\lambda}-\frac{l^2_{0}}{r^2}-\Omega^2\,r^2-b\,r \right]\,\psi (r)=0.
\label{41}
\end{equation} 

Introduce a new variable $\rho=\sqrt{\Omega}\,r$, then the Eq. (\ref{41}) becomes
\begin{equation}
\left [\frac{d^2}{d\rho^2}+\frac{1}{\rho}\,\frac{d}{d\rho}+\frac{\tilde{\lambda}}{\Omega}-\frac{l^2_{0}}{\rho^2}-\rho^2-\theta\,\rho \right]\,\psi (\rho)=0.
\label{42}
\end{equation}

Let the possible solution to Ee. (\ref{42}) is 
\begin{equation}
\psi=\rho^{|l_0|}\,e^{-\frac{1}{2}\,(\theta+\rho)\,\rho}\,H (\rho)
\label{solution}
\end{equation}

Substituting Eq. (\ref{solution}) into the Eq. (\ref{42}), we obtain
\begin{equation}
H''(\rho)+\left [\frac{(1+2\,|l_0|)}{\rho}-\theta-2\,\rho \right ]\,H'(\rho)+\left [-\frac{\frac{\theta}{2}\,(1+2\,|l_0|)}{\rho}+\Theta_0 \right]\,H (\rho)=0,
\label{43}
\end{equation}
where $\Theta_0=\frac{\tilde{\lambda}}{\Omega}-2\,(1+|l_0|)+\frac{\theta^2}{4}$. 

Equation (\ref{43}) is the biconfluent Heun's differential equation \cite{IJMPD,AHEP,EPJP,EPJC,AHEP2,AR,SYS, ERFM,KBCF2} and $H (\rho)$ is the Heun polynomials.

Substituting the power series solution (\ref{17}) into the Eq. (\ref{43}), we obtain the following recurrence relation for the coefficients:
\begin{equation}
c_{n+2}=\frac{1}{(n+2)(n+2+2\,l_0)}\,\left[\frac{\theta}{2}\,(2\,n+3+2\,|l_0|)\,c_{n+1}-(\Theta_0-2\,n)\,c_{n} \right].
\label{44}
\end{equation}
And the various coefficients are
\begin{eqnarray}
&&c_1=\frac{\theta}{2}\,c_0,\nonumber\\
&&c_2=\frac{1}{4\,(1+j)}\,\left [\frac{\theta}{2}\,(3+2\,|l_0|)\,c_{1}-\Theta_0\,c_{0} \right].
\label{45}
\end{eqnarray}.

The power series expansion (\ref{17}) becomes a polynomial of degree $n$ by imposing two conditions \cite{IJMPD,AHEP,EVBL,EVBL2,EVBL3,EPJP,EPJC,AHEP2,ERFM,KBCF2}:
\begin{equation}
c_{n+1}=0\quad,\quad \Theta_0=2\,n \quad (n=1,2,3,4,....) 
\label{46}
\end{equation}

By analyzing the condition $\Theta_0=2\,n$, we get the following energy eigenvalues $E_{n,l}$:
\begin{equation}
E_{n,l}=\pm\,\sqrt{k^2+q^2+m^2+2\,m\,\omega\,l_0+2\,\Omega\,\left(n+1+\frac{|l-\frac{q\,\Phi}{2\pi}|}{\alpha}\right)-\frac{m^2\,\eta^2_{L}}{\Omega^2}},
\label{47}
\end{equation}
where $l_0=\frac{1}{\alpha}\,(l-\frac{q\,\Phi}{2\,\pi})$.

For the radial mode $n=1$, $c_2=0$ which implies
\begin{equation}
\Omega_{1,l}=\left [\frac{m^2\,\eta^2_{L}}{2}\,(3+2\,|l_0|) \right]^{\frac{1}{3}}.
\label{48}
\end{equation}
a constraint on the parameter $\Omega_{1,l}$. Therefore, the magnetic field is given by
\begin{eqnarray}
\omega_{1,l}&=&\frac{1}{m}\,\sqrt{\Omega^2_{1,l}-\eta^2_{L}}\nonumber\\\Rightarrow
B^{1,l}_{0}&=&\frac{2}{q}\,\sqrt{\Omega^2_{1,l}-\eta^2_{L}}\nonumber\\
&=&\frac{2}{q}\,\sqrt{ \left [\frac{m^2\,\eta^2_{L}}{2}\,(3+2\,|l_0|) \right]^{\frac{2}{3}}-\eta^2_{L}}.
\label{49}
\end{eqnarray}

Therefore the ground state energy levels 
\begin{eqnarray}
E_{1,l}&=&\pm\,\{k^2+q^2+m^2+2\,m\,\omega_{1,l}\,\frac{(l-\frac{q\,\Phi}{2\pi})}{\alpha}\nonumber\\
&&+2\,\Omega_{1,l}\,\left(n+1+\frac{|l-\frac{q\,\Phi}{2\pi}|}{\alpha}\right)-\frac{m^2\,\eta^2_{L}}{\Omega^2_{1,l}}\}^{\frac{1}{2}}.
\label{50}
\end{eqnarray}

Equation (\ref{50}) corresponds to the allowed values of relativistic energy level for the radial mode $n=1$ of the system subject to linear confining potential in the background of Kaluza-Klein theory. For $\alpha \rightarrow 1$, the energy eigenvalues is consistent with those result in \cite{EVBL2}. Since the values of the wedge parameter $\alpha$ lies in the range $\alpha <1$, hence, the energy levels presented here shifted in comparison to those in \cite{EVBL2} and break their degeneracy. We observed here that the magnetic field $B_0$ depends on the quantum numbers with their possible values are determined by a relation. The energy level get shifted due to the presence of quantum flux $\Phi$ in the system which gives us an analogous of the Aharonov-Bohm effect making them a periodic function with periodicity $\Phi_0=\frac{2\,\pi}{q}\,\tau$, with $\tau=0,1,2,...$, that is, $E_{n, l} (\Phi+\Phi_0)=E_{n, l \mp \tau} (\Phi)$.

The ground state wave-function ($n=1$) is given by
\begin{equation}
    \psi_{1,l}=\rho^{|l_0|}\,e^{-\frac{1}{2}\,(2\,\sqrt{\frac{2}{3+2\,|l_0|}}+\rho)\,\rho}\,(c_0+c_1\,\rho),
    \label{51}
\end{equation}
where
\begin{equation}
    c_1=\sqrt{\frac{2}{3+2\,|l_0|}}\,c_0.
    \label{52}
\end{equation}

\vspace{0.5cm}
{\bf Special case}
\vspace{0.5cm}

In this special case, we choose zero magnetic field, $B_0 \rightarrow 0$ in the above system. The radial wave-equations Eq. (\ref{41}) becomes
\begin{equation}
\left [\frac{d^2}{dr^2}+\frac{1}{r}\,\frac{d}{dr}+\tilde{\lambda}-\frac{l^2_{0}}{r^2}-\eta^2_{L}\,r^2-b\,r \right]\,\psi (r)=0.
\label{cc1}
\end{equation}
Transforming $\rho=\sqrt{\eta_L}\,r$ into the Eq. (\ref{41}), we have
\begin{equation}
\left [\frac{d^2}{d\rho^2}+\frac{1}{\rho}\,\frac{d}{d\rho}+\frac{\tilde{\lambda}}{\eta_L}-\frac{l^2_{0}}{\rho^2}-\rho^2-\frac{b}{\eta^{\frac{3}{2}}_L}\,\rho \right]\,\psi (r)=0.
\label{cc2}
\end{equation} 

Let us now discuss the asymptotic behavior of the possible solutions to Eq. (\ref{cc2}), that is, we hope that $\psi (\rho) \rightarrow 0$ at $\rho \rightarrow 0$ and $ \rho \rightarrow \infty$. Suppose the possible solution to Eq. (\ref{cc2}) is
\begin{equation}
\psi (\rho)=\rho^{|l_0|}\,e^{-\frac{1}{2}\,(\rho+\frac{b}{\eta^{\frac{3}{2}}_L})\,\rho}\, H (\rho)
\label{cc3}
\end{equation}
Substituting the solution Eq. (\ref{cc3}) into the Eq. (\ref{cc2}), we obtain
\begin{eqnarray}
&&H '' (\rho)+\left [\frac{1+2\,|l_0|}{\rho}-2\,\rho-\frac{b}{\eta^{\frac{3}{2}}_L} \right]\,H' (\rho)\nonumber\\
&&+\left [\frac{\tilde{\lambda}}{\eta_L}+\frac{b^2}{4\,\eta^{3}_L}-2\,(1+|l_0|)-\frac{\frac{b}{2\,\eta^{\frac{3}{2}}_L}\,(1+2\,|l_0|)}{\rho} \right]\,H (\rho)=0.
\label{cc4}
\end{eqnarray}

Substituting the power series solution Eq. (\ref{14}) into the above equation, we get 
\begin{eqnarray}
c_{n+2}&=&\frac{1}{(n+2)(n+2+2\,|l_0|)}\,[\frac{b}{\eta^{\frac{3}{2}}_L}\,(n+\frac{3}{2}+|l_0|)\,c_{n+1}\nonumber\\
&&-\left\{\frac{\tilde{\lambda}}{\eta_L}+\frac{b^2}{4\,\eta^{3}_L}-2\,(1+|l_0|)-2\,n \right\}\,c_n],
\label{cc5}
\end{eqnarray}
where few coefficients are
\begin{eqnarray}
c_1&=&\frac{b}{2\,\eta^{\frac{3}{2}}_L}\,c_0,\nonumber\\
c_2&=&\frac{1}{4\,(1+|l_0|)}\,\left [\frac{b}{2\,\eta^{\frac{3}{2}}_L}\,(3+2\,|l_0|)\,c_1-\left (\frac{\tilde{\lambda}}{\eta_L}+\frac{b^2}{4\,\eta_L}-2-2\,|l_0|\right)\,c_0 \right].
\label{cc6}
\end{eqnarray}

The power series solution becomes a polynomial of degree $n$. for this, we must have \cite{IJMPD,AHEP,EVBL,EVBL2,EVBL3,EPJP,EPJC,AHEP2,ERFM,KBCF2}
\begin{equation}
\frac{\tilde{\lambda}}{\eta_L}+\frac{b^2}{4\,\eta^{3}_L}-2\,(1+|l_0|)=2\,n\quad (n=1,2,...),\quad c_{n+1}=0.
\label{cc7}
\end{equation}

For $n=1$, we have $c_2=0$ which implies from (\ref{cc6})
\begin{equation}
\eta_{1,l\,L}=\frac{m^2}{2}\,(3+2\,|l_0|).
\label{cc8}
\end{equation}
a constraint on the potential parameter $\eta_{1,l}$.

By analysing the condition $\frac{\tilde{\lambda}}{\eta_L}+\frac{b^2}{4\,\eta_L}-2\,(1+|l_0|)=2\,n$, we get
\begin{equation}
E_{n,l}=\pm\,\sqrt{k^2+q^2+2\,\eta_L\,(n+1+|l_0|)}.
\label{cc9}
\end{equation}
Therefore, the ground state energy eigenvalue is given by
\begin{eqnarray}
E_{1,l}&=&\pm\,\sqrt{k^2+q^2+2\,\eta_{1,l\,L}\,(2+|l_0|)}\nonumber\\
&=&\pm\,m\,\sqrt{\frac{k^2}{m^2}+\frac{q^2}{m^2}+(3+2\,\frac{|l-\frac{q\,\Phi}{2\,\pi}|}{\alpha})\,(2+\frac{|l-\frac{q\,\Phi}{2\,\pi}|}{\alpha})}.
\label{cc10}
\end{eqnarray}

Equation (\ref{cc10}) represents the energy level associated with radial mode $n=1$ of scalar particle in background governed by Kaluza-Klein theory without a magnetic field. For $\alpha \rightarrow 1$, the energy eigenvalue is consistent with those result in \cite{EVBL3}. Since the values of the wedge parameter $\alpha$ lies in the range $\alpha <1$, hence, the energy levels presented here shifted in comparison to those in \cite{EVBL3} and break their degeneracy. The energy level get shifted due to the presence of quantum flux $\Phi$ in the system which gives us an analogous of the Aharonov-Bohm effect making them a periodic function with periodicity $\Phi_0=\frac{2\,\pi}{q}\,\tau$, with $\tau=0,1,2,...$, that is, $E_{n, l} (\Phi+\Phi_0)=E_{n, l \mp \tau} (\Phi)$.

\vspace{0.5cm}
{\bf Case D\quad:\quad Without scalar potential}
\vspace{0.5cm}

In this case, we consider zero scalar potential, $S=0$, in the considered system discussed in section 2. Therefore, the radial wave-equations Eq. (\ref{8}) becomes
\begin{equation}
\left [\frac{d^2}{dr^2}+\frac{1}{r}\,\frac{d}{dr}+\lambda_0-\frac{l^2_{0}}{r^2}-m^2\,\omega^2\,r^2 \right]\,\psi (r)=0,
\label{dd1}
\end{equation}
where
\begin{eqnarray}
\lambda_0&=&E^2-k^2-q^2-m^2-2\,m\,\omega\,\frac{(l-\frac{q\,\Phi}{2\pi})}{\alpha},\nonumber\\
\omega&=&\frac{q\,B_0}{2\,m}.
\label{dd2}
\end{eqnarray}

Transforming to a new variable $\rho=m\,\omega\,r^2$ into the Eq. (\ref{dd1}), we obtain \cite{IJMPA,AFN}
\begin{equation}
\psi '' (\rho)+ \frac{1}{\rho}\,\psi' (\rho)+\frac{1}{\rho^2}\,\left (-\xi_1\,\rho^2+\xi_2\,\rho-\xi_3 \right)\,\psi (\rho)=0,
\label{dd3}
\end{equation}
where 
\begin{equation}
\xi_1=\frac{1}{4}\quad,\quad \xi_2=\frac{\lambda_0}{4\,m\,\omega}\quad,\quad \xi_3=\frac{l^2_{0}}{4}.
\label{dd4}
\end{equation}

Therefore, the energy eigenvalues is given by
\begin{equation}
E_{n',l}=\pm\,\sqrt{k^2+q^2+m^2+q\,B_0\,\left (2\,n'+1+\frac{(l-\frac{q\,\Phi}{2\pi})}{\alpha}+\frac{|l-\frac{q\,\Phi}{2\pi}|}{\alpha} \right)},
\label{dd5}
\end{equation}
where $n'=0,1,2,...$.

Equation (\ref{dd5}) is the closed expression of the energy eigenvalues of a scalar particle subject to a uniform magnetic field including magnetic quantum flux in $5D$ cosmic string space-time geometry within the Kaluza-Klein theory. For zero magnetic quantum flux, $\Phi \rightarrow 0$, the energy eigenvalues Eq. (\ref{dd5}) is consistent with those result in \cite{CF}. Thus we can see that the energy eigenvalues Eq. (\ref{dd5}) get modify in comparison to those in \cite{CF} due to the presence of magnetic quantum flux $\Phi$ which gives rise to a relativistic analogue of the Aharonov-Bohm effect. The energy level get shifted due to the presence of quantum flux $\Phi$ in the system which gives us an analogous of the Aharonov-Bohm effect making them a periodic function with periodicity $\Phi_0=\frac{2\,\pi}{q}\,\tau$, with $\tau=0,1,2,...$, that is, $E_{n', l} (\Phi+\Phi_0)=E_{n', l \mp \tau} (\Phi)$.

The wave-functions is given by
\begin{equation}
\psi_{n',l} (\rho)=|N|_{n',l}\,\rho^{\frac{|l-\frac{q\,\Phi}{2\pi}|}{2\,\alpha}}\,e^{-\frac{\rho}{2}}\,L^{(\frac{l-\frac{q\,\Phi}{2\pi}}{\alpha})}_{n'} (\rho),
\label{dd6}
\end{equation}
where $|N|_{n',l}=\left(\frac{n'!}{2\,\left (n'+\frac{|l-\frac{q\,\Phi}{2\pi}|}{\alpha} \right)!}\right)^{\frac{1}{2}}$ is the normalization constant and $L^{(\frac{l-\frac{q\,\Phi}{2\pi}}{\alpha})}_{n'} (\rho)$ is the generalized Laguerre polynomials.

\section{Conclusions}

In this work, we have investigated the relativistic quantum dynamics of spin-$0$ scalar particle in the presence of a uniform magnetic field and magnetic quantum flux which is introduced into the system through an extra dimension of the cosmic string space-time under the effects of scalar potential. In this scenario, we have analyzed the interaction between a scalar particle and scalar potential, where, in the search for solutions of bound states, we determine analytically the energy eigenvalue of the system, which is influenced by the background and the wedge parameter. We can note that relativistic energy spectrum is not defined by a closed expression. In fact, it is only possible to determine allowed values of the relativistic energy levels for each radial mode separately. As an example, we analyze the lowest energy state of the system represented by the radial mode $n=1$, instead of $n=0$. 

In {\bf Case A}, we have analyzed interaction between a scalar particle and Cornell-type scalar potential in the considered relativistic system in the background governed by the KKT. We have solved the radial wave equation of the Klein-Gordon equation and reduce to the Biconfluent Heun's differential equation. By power series method, we have solved this equation and obtained the non-closed expression of the energy profile (\ref{22}) of the system. By imposing the recurrence relation $c_{n+1}=0$, we have obtained the lowest energy state (\ref{25}) as well as eigenfunction (\ref{26})-(\ref{27}). We have seen that the presence of a Cornell-type scalar potential as well as the wedge parameter breaks the degeneracy of the Landau levels. In addition, a quantum effect characterized by the dependence the magnetic field on the quantum numbers of the system is observed, where we have shown that their possible values are determined by a third-degree algebraic equation. 

In {\bf Case B}, we have considered a Coulomb-type scalar potential and obtained the non-closed expression of the energy eigenvalues (\ref{37}). By imposing the recurrence relation $c_{n+1}=0$, we have obtained the lowest energy state (\ref{400}) as well as the eigenfunction (\ref{410}) of the system. We have seen that the presence of a Coulomb-type scalar potential as well as the wedge parameter breaks the degeneracy of the Landau levels. We have observed a quantum effect due to the dependence of the magnetic field on the quantum numbers of the system which we determined by a relation for the different radial mode $n=1,2,....$. Furthermore, we have discussed a special case corresponds to zero external magnetic field, $B_0 \rightarrow 0$, in the considered relativistic system. In that case, the radial wave-equation can be solved using the Nikiforov-Uvarof method \cite{AFN} and obtained the energy eigenvalues (\ref{bb4}) and the radial wave-function (\ref{bb5}). We have seen that the presence of the wedge parameter shifted the energy levels in comparison to those in \cite{EVBL} and break their degeneracy.

In {\bf Case C}, we have considered a linear confining potential and obtained the non-closed expression of the energy eigenvalues (\ref{47}). By imposing the recurrence relation $c_{n+1}=0$, we have obtained the lowest energy state (\ref{50}) as well as the eigenfunction (\ref{51})--(\ref{52}) of the system. We have seen that the presence of a linear confining potential as well as the wedge parameter breaks the degeneracy of the Landau levels. We have observed a quantum effect due to the dependence of the magnetic field on the quantum numbers of the system which we determined by a relation for the different radial mode $n=1,2,....$. Furthermore, we have discussed a special case corresponds to zero external magnetic field, $B_0 \rightarrow 0$, in the considered relativistic system and obtained the energy eigenvalues (\ref{cc9}). Following the similar technique we have obtained the lowest energy state and wave-function for the radial mode $n=1$. In this case, we have observed that the presence of the wedge parameter, $\alpha <1$, modify the energy spectrum of the quantum system and shifted the energy level in comparison to those \cite{EVBL2} and \cite{EVBL3}, respectively.

In {\bf Case D}, we have solved the Klein-Gordon equation subject to a uniform magnetic field including  magnetic quantum flux in the background of Kaluza-Klein theory without any potential. We have solved the radial wave-equation using the Nikiforov-Uvarof method \cite{AFN} and obtained the energy eigenvalues (\ref{dd5}) and the radial wave-function (\ref{dd6}). We have seen that the presence of the magnetic quantum flux $\Phi$ modified the energy eigenvalues of the relativistic system in comparison to those in \cite{CF}.

In all cases, we have observed that the angular momentum number $l$ is shifted, $l \rightarrow l_0=\frac{1}{\alpha}\,(l-\frac{q\,\Phi}{2\,\pi})$, an effective angular quantum number. Therefore, the relativistic energy eigenvalues depends on the geometric quantum phase \cite{YA,GAM}. Thus, we have that, $E_{n, l} (\Phi+\Phi_0)=E_{n, l \mp \tau} (\Phi)$, where $\Phi_0=\pm\,\frac{2\pi}{q}\,\tau$, with $\tau=0,1,2,...$. This dependence of the relativistic energy eigenvalues on the geometric quantum phase gives rise to a relativistic analogue of the Aharonov-Bohm effect \cite{MP,VBB,YA}.

\end{document}